\documentclass[a4paper, secnumarabic, twocolumn, amssymb, nobibnotes, aps]{revtex4-1}

\usepackage{graphicx}
\usepackage{pslatex}
\usepackage{xspace}
\usepackage{amsmath}
\usepackage{color}
\usepackage{booktabs}
\usepackage{hyperref}
\begin{document}

\title{Impact of nongeminate recombination on the performance of pristine and annealed P3HT:PCBM solar cells} 

\author{M.~Gluecker$^1$}
\author{A.~Foertig$^1$}\email{afoertig@physik.uni-wuerzburg.de}
\author{V.~Dyakonov$^{1,2}$}
\author{C.~Deibel$^1$}
\affiliation{$^1$ Experimental Physics VI, Julius-Maximilians-University of W\"urzburg, D-97074 W\"urzburg, Germany}
\affiliation{$^2$ Bavarian Center for Applied Energy Research e.V. (ZAE Bayern), D-97074 W\"urzburg, Germany}

\begin{abstract}
Transient photovoltage (TPV) and voltage dependent charge extraction (CE) measurements were applied to poly(3-hexylthiophene)(P3HT):[6,6]-phenyl-C61 butyric acid methyl ester (PCBM) bulk heterojunction solar cells to analyze the limitations of solar cell performance in pristine and annealed devices. From the determined charge carrier decay rate under open circuit conditions and the voltage dependent charge carrier densities $n(V)$ the nongeminate loss current $j_{loss}$ of the device is accessible. We found that $j_{loss}$ alone is sufficient to describe the j-V characteristics across the whole operational range, for annealed and, not yet shown before, also for the lower performing pristine solar cells. Even in a temperature range from 300~K to 200~K nongeminate recombination is found to be the dominant and, therefore, performance limiting loss process. Consequently, charge photogeneration is voltage independent in the voltage range studied.
\end{abstract}

This is the pre-peer reviewed version of the following article: Impact of nongeminate recombination on the performance of pristine and annealed P3HT:PCBM solar cells, M. Gluecker, A. Foertig, V. Dyakonov, C. Deibel, which has been published in final form at {\bf \href{http://onlinelibrary.wiley.com/doi/10.1002/pssr.201206248/abstract}{Rapid\ Research\ Letters. (DOI: 10.1002/pssr.201206248)}}

\maketitle   % please do not remove

%\section{Introduction}

Organic solar cells based on polymer--fullerene blend systems are considered as a lower cost alternative to their inorganic counterparts. But even though a significant progress has been made during the last years, recent power conversion efficiencies up to 9.2\%~\cite{science2011} are still too low for profitable applications.
One key challenge to further improve the solar cells is to understand the performance limiting loss mechanisms. Already separated polarons on their way towards the respective electrodes can be lost due to nongeminate recombination processes~\cite{deibel2010}.
In this paper we show for temperatures from room temperature to 200K that nongeminate recombination is the predominant loss process determining the shape of the current--voltage (j-V) characteristics of annealed as well as pristine (as-prepared) P3HT:PCBM solar cells and, therefore, is the limiting factor in terms of photovoltaic device performance.
%\section{Background}

In this study TPV~\cite{shuttle2008b,foertig2009} and CE~\cite{shuttle2008a,shuttle2010} were used, which allow us to determine a small perturbation charge carrier lifetime $\tau_{\Delta n}(n)$ and voltage dependent charge carrier densities $n(V)$, respectively. For further analysis we utilised an approach introduced by Shuttle et al.~\cite{shuttle2010}. The current density $j$ is defined by the continuity equation, $\frac{1}{q}\frac{dj}{dx}+G-R=0,$ shown here for steady state conditions, with the photogeneration rate $G$, the recombination rate $R$, the spatial derivative of the current $j$ and the elementary charge $q$.
By integrating the continuity equation we are able to divide the current into a generation term $j_{gen}$ and a loss term $j_{loss}$,
\begin{equation}
\label{eq:cont_eq_int}
j=qdG-qdR=j_{gen}-j_{loss},
\end{equation}
with the active layer thickness $d$. Thereby we neglect explicit effects of charge injection as we focussed on the voltage range between 0 V and $V_{oc}$. 
The small perturbation charge carrier lifetime $\tau_{\Delta n}$ revealed by TPV is related to the total charge carrier lifetime $\tau (n)$ by
\begin{equation}
\label{eq:tau}
\tau (n)=\tau_{\Delta n}\cdot (\lambda+1),
\end{equation}
where $\lambda$+1 is the apparent recombination order according to $R\propto n^{\lambda+1}$~\cite{hamilton2010,foertig2009}.
\begin{figure}[b]
\begin{center}
  \includegraphics*[width=.34\textwidth]{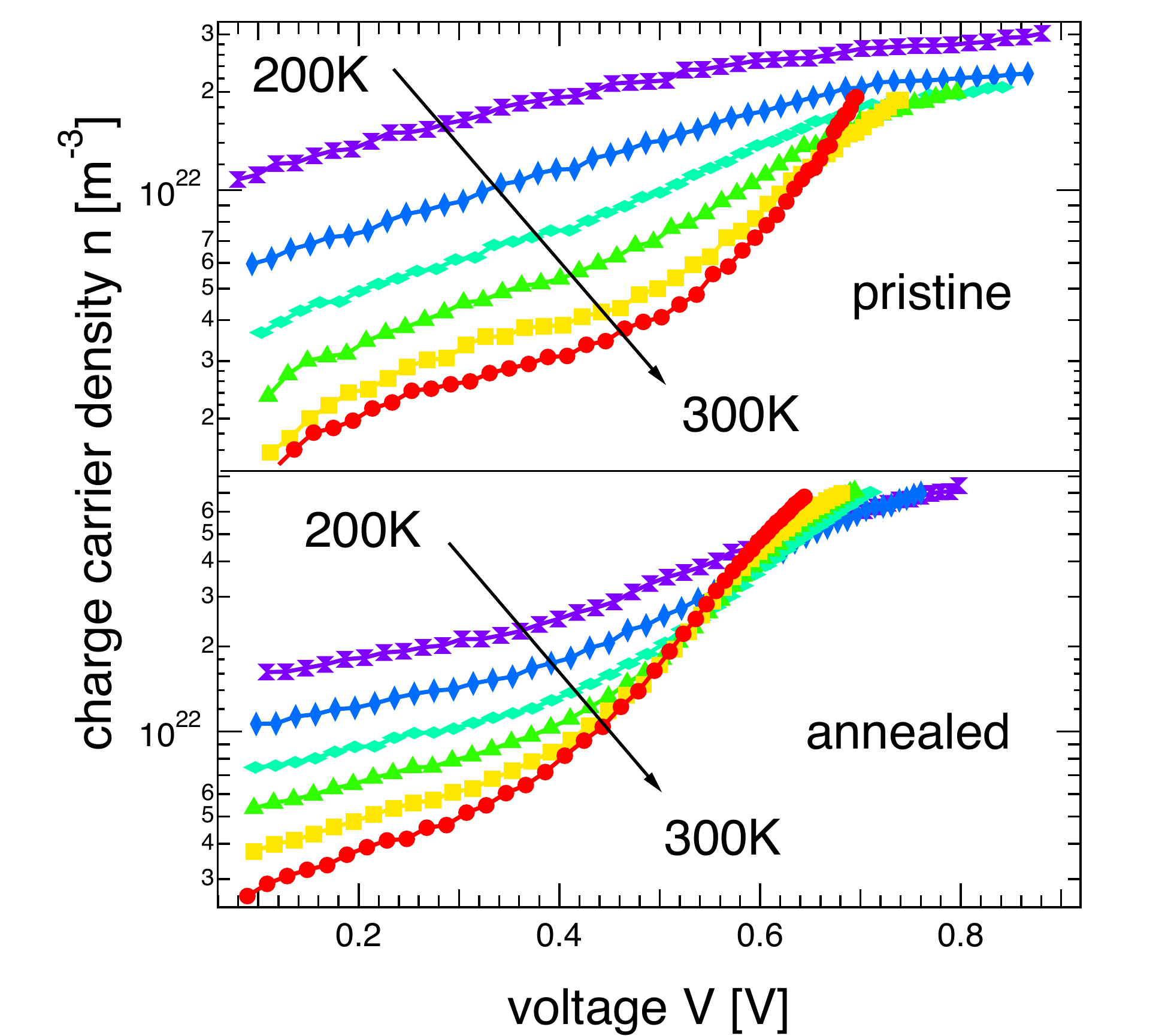}
  \caption{ Charge carrier density $n$, as a function of the voltage $V$ across the device, obtained by CE experiments for different temperatures in the range of 300K to 200K at a bias light equivalent to 1 sun.}
    \label{fig:n_V}
    \end{center}
\end{figure}
Experimentally, the nongeminate recombination rate can be calculated by $R$=$\frac{n}{\tau(n)}$,
and enables to determine the nongeminate loss current $j_{loss}=qdR$. Spatial variations of the charge carrier density $n$ were not considered, as they cannot be directly determined by extraction experiments. 
We assumed negligible nongeminate losses under short circuit conditions and voltage independent charge photogeneration. 
Thus, a constant generation current $j_{gen}=-qdG\approx j_{sc}$ was defined for all considered voltages. Accordingly, the overall device current density was approximated by
\begin{equation}
	\label{eq:J_V}
	j(V)\approx j_{sc}-qd\frac{n(V)}{\tau (n(V))}.
\end{equation}
%\section{Experimental methods}

The P3HT:PCBM devices considered were prepared as described in detail in Ref.~\cite{rauh2012}, with layer thicknesses between 200-230~nm. Samples referred to as annealed were heated for 10~minutes at 130$^{\circ}$C before contact evaporation.

An Oriel 1160 AM1.5G solar simulator was used in inert atmosphere to perform illuminated current--voltage measurements. A power conversion efficiency of 3.2\% was achieved for the annealed and 2.0\% for the pristine device.
For the T-dependent TPV and CE measurements the samples were transferred into a closed-cycle optical cryostat with helium as contact gas. TPV measurements were performed using a Nd:YAG laser ($\lambda$=532 nm, pulse duration $<$ 80 ps).  For the charge extraction studies a 10 W white light LED (Seoul) including focusing optics,  and a MOSFET transistor as driver were used. TPV and CE transients were acquired by a digital storage oscilloscope (Agilent Infiniium DS090254A) and the corresponding voltages $V_{app}$ were applied by a Keithley 2602 source meter together with a fast analog switch (response time $\approx$ 50~ns).

%\section{Results and discussion}

Fig. \ref{fig:n_V} shows the charge carrier density $n$ as a function of the voltage $V=V_{app}-R_s I$ across the cell, with a series resistance of $R_s=92 \Omega$ for the annealed and $R_s=217 \Omega$ for the pristine device, respectively. Both were derived from the ohmic range of the dark j-V curve at 300K. The total extractable charge obtained from CE was corrected for capacitive charges by performing CE measurements under reverse bias in the dark. Incurred recombination losses during extraction were considered by an iteration procedure similar to the one introduced in Ref.~\cite{shuttle2008a}.
\begin{figure}[bt]
\begin{center}
  \includegraphics*[width=.34\textwidth]{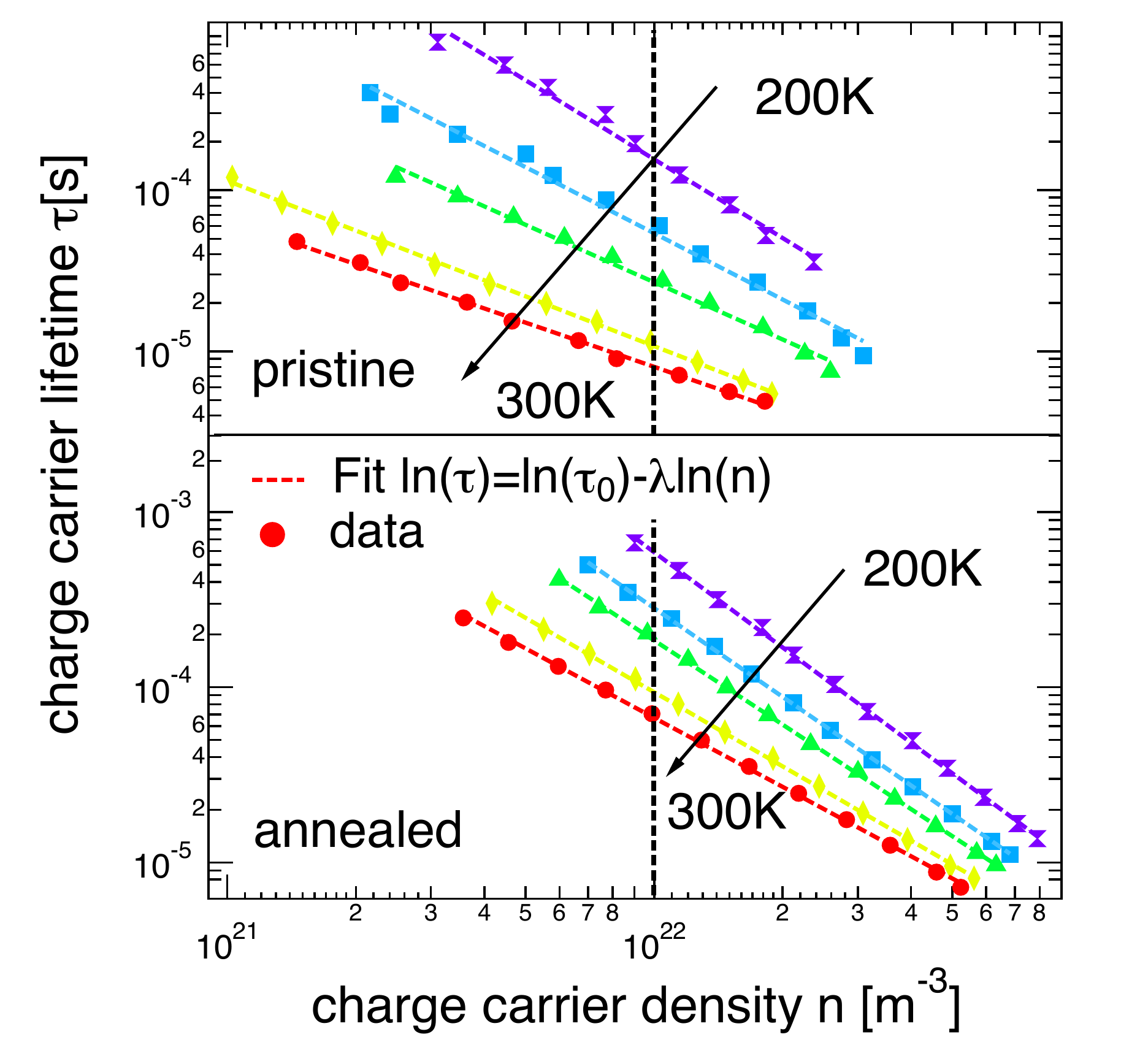}
  \caption{Comparison of the charge carrier lifetime $\tau$ derived by Eq.~(\ref{eq:tau}) as a function of the charge carrier density $n$, and respective fits (dashed lines).}
    \label{fig:tau_n}
    \end{center}
\end{figure}
The charge carrier densities of the annealed sample are higher than the appropriate pristine values (Fig.~\ref{fig:n_V}). This is consistent with a better exciton harvesting and charge collection due to an increased degree of crystallinity and an improved absorption \cite{chirvase2004,vanlaeke2006}, as well as longer charge carrier lifetimes (Fig.~\ref{fig:tau_n}) in annealed cells.%$R$ was calculated from Eq.~(\ref{eq:R}) by using $n(V_{oc},T)$ data determined by CE as well as the respective charge carrier lifetime $\tau$ under open circuit conditions from TPV.%

Furthermore Fig.~\ref{fig:n_V} reveals that for both samples the average carrier concentration $n$ increases for lower temperatures, while the dependence of $n$ on $V$ becomes weaker, especially at higher voltages. The former can be attributed to a reduction of the charge carrier mobility and thus to a decrease of the recombination rate $R$. This leads to an increase in charge carrier lifetime $\tau$ (Fig.~\ref{fig:tau_n}). The reduced slope of $n (V)$ at lower temperatures, especially for the pristine cell, could result from a space charge limitation, hindering charge injection at a certain applied voltage as well as extraction in CE experiments. 

As depicted by the slopes of the fits in Fig.~\ref{fig:tau_n} the carrier concentration dependence of $\tau(n)=\tau_{0}{n}^{-\lambda}$ becomes stronger at low temperatures, %, discussed in literature with respect to an enhanced dependence of charge carrier mobility on charge carrier density~\cite{Mauer2010}.
where $\tau_{0}$ corresponds to the intercept at $n=0$ \cite{maurano2010}. Thus, the charge decay order $\lambda+1$, as defined above, is accessible by the slope of $\tau(n)$~\cite{shuttle2008b,hamilton2010}. Its magnitude increases at lower temperatures, which indicates additional trapped and subsequently released charges being involved in recombination processes, as reported previously for annealed devices~\cite{foertig2009,rauh2012}. In general $\tau$ is higher for the annealed than for the pristine cell at  matched $n$ and temperature. This finding can result from distinct phase separation and crystalline domains within the bulk due to annealing~\cite{vanbavel2010}. Absolute values for 300K are in good agreement with Ref.~\cite{hamilton2010}.

\begin{figure}[bt]
\begin{center}
\includegraphics*[width=.40\textwidth]{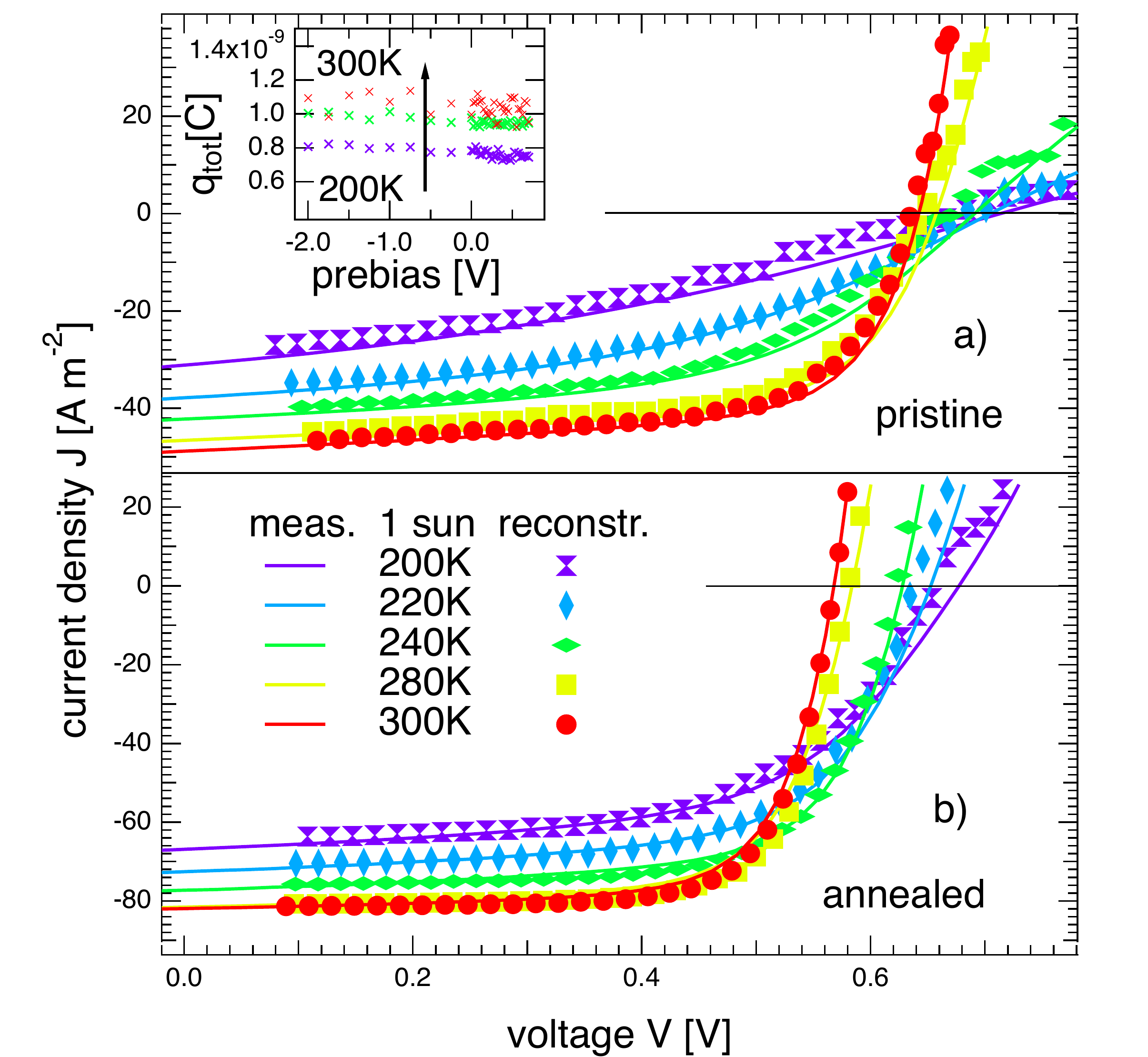}
\caption{T-dependent j-V curves for an annealed (a) and pristine (b) P3HT:PCBM device (bold lines) for 1 sun, compared with the reconstructed j-V curves (symbols), derived from TPV and CE measurements using Eq. (\ref{eq:J_V}). Inset: TDCF data for the pristine sample. $q_{tot}$ marks the total amount of extracted charges after photoexcitation 
for different applied prebias voltages.}
\label{fig:reconstruction}
\end{center}
\end{figure}

Employing Eq.~(\ref{eq:J_V}) and the experimental TPV and CE data described above, we were able to reconstruct the j-V behaviour of the devices. In Fig.~\ref{fig:reconstruction}, the directly measured j-V characteristics are shown together with the reconstructed ones at 1~sun illumination. All voltages were corrected for the series resistance $R_s$. The fill factor and the open circuit voltage could be reproduced quite accurately for both cells. 

As the recombination rate $R$ and, thus, $j_{loss}$ are determined by charge carrier densities derived from charge extraction measurements, only nongeminate recombination losses are included, since only separated, free charge carriers can be extracted during these experiments.
Hence, the good agreement of measured and reconstructed data confirms that nongeminate recombination is the main loss mechanism determining the j-V behaviour for annealed \cite{shuttle2010} as well as for pristine P3HT:PCBM devices.
We point out that these findings imply a voltage independent photogeneration down to 200K. Similar results concerning photogeneration were found recently for annealed samples \cite{street2010,kniepert2011,mingebach2012}, but were not yet reported for pristine ones. We verified the constant photogeneration rate of the pristine device by time delayed collection field (TDCF) measurements (inset Fig. \ref{fig:reconstruction})
~\cite{mingebach2012}.  A field dependence less than 5\% is observable between open and short circuit conditions, negligible compared to the experimental error. 

%\section{Conclusion}
To conclude, transient photovoltage and charge extraction measurements were employed to determine the charge carrier lifetime and density in pristine and annealed P3HT:PCBM bulk heterojunction solar cells. Illumination density, bias voltage and temperature were varied. A loss current representing nongeminate recombination was calculated over the entire operation regime of the photovoltaic devices. These losses, in combination with a voltage independent charge photogeneration term, were used to reconstruct the j-V characteristics of the solar cells. The good agreement between measured and reconstructed j-V characteristics clearly reveals that nongeminate charge recombination is the performance limiting loss mechanism in P3HT:PCBM solar cells. This further implies that the assumption of voltage independent charge generation holds for both, pristine and annealed devices, for the temperatures considered here.

\begin{acknowledgements}
The current work is supported by the Bundesministerium f{\"u}r Bildung und Forschung in the framework of the Grekos Project (Contract No. 03SF0356B) and by the Bavarian Ministry of Economic Affairs, Infrastructure, Transport and Technology.
\end{acknowledgements}

\end{document}